\documentstyle[sprocl]{article}

\def\PRD{{\em Phys. Rev.} D}
\def\ZPC{{\em Z. Phys.} C}

\def\be{\begin{equation}}
\def\ee{\end{equation}}
\def\bea{\begin{eqnarray}}
\def\eea{\end{eqnarray}}

\begin{document}

\begin{flushright}                                
UMN-D-00-1 \\ February 2000 \\ \vspace{0.3in}
\end{flushright}

\title{PAULI-VILLARS REGULARIZATION AND DISCRETE LIGHT-CONE
QUANTIZATION IN YUKAWA THEORY%
\footnote{To appear in the proceedings of               
the CSSM Workshop on Light-Cone QCD
and Nonperturbative Hadron Physics,
Adelaide, Australia, December 13-21, 1999.}%
}

\author{J. R. Hiller%
\footnote{\baselineskip=14pt                            
Work supported in part by the Department of Energy,
contract DE-FG02-98ER41087.}%
}

\address{%
Department of Physics, 
University of Minnesota Duluth \\ 
Duluth, MN 55812, USA \\
E-mail: jhiller@d.umn.edu} 

\maketitle
\abstracts{%
The techniques of Pauli--Villars regularization and
discrete light-cone quantization are combined to
analyze Yukawa theory in a single-fermion truncation.
A special form of the Lanczos algorithm is constructed
for diagonalization of the indefinite-metric
light-cone Hamiltonian.}

\section{Introduction}

As a step toward development of a method for nonperturbative
solution of four-dimensional quantum field theories, 
we consider a single-fermion truncation of Yukawa
theory in discrete light-cone quantization (DLCQ).~\cite{DLCQreview}
To regulate the theory we introduce Pauli-Villars (PV)
bosons with indefinite metric~\cite{PauliVillars} into
the Fock basis.  This extends earlier work on model
theories by Brodsky, Hiller and McCartor~\cite{PV1,PV2} to
a more physical situation.

The importance of Pauli--Villars regularization stems from
its ability to regulate the continuum theory with 
counterterms generated automatically, except for a trivial
mass counterterm.  The numerical method, in this case DLCQ,
can then be applied to a finite theory.  The bare parameters
are fixed via physical constraints and become
functions of the PV masses and of any numerical parameters.
The original theory is recovered in the following 
sequence of limits: infinite numerical resolution, infinite
(momentum) volume, and infinite PV masses.

The DLCQ formulation is based on
periodic boundary conditions for bosons and antiperiodic 
conditions for fermions
in a light-cone box $-L<x^-\equiv(t-z)<L$, $-L_\perp<x,y<L_\perp$.
The light-cone 3-momentum $\underline{p}\equiv(p^+,\vec{p}_\perp)$
is then on a discrete grid specified by integers 
$\underline{n}=(n,n_x,n_y)$: 
$p^+\equiv(E+p_z)=\frac{\pi}{L}n$,
$\vec{p}_\perp=(\frac{\pi}{L_\perp}n_x,\frac{\pi}{L_\perp}n_y)$.
The limit $L\rightarrow\infty$ can be exchanged for a limit
in terms of the integer {\em resolution}~\cite{PauliBrodsky}
$K\equiv\frac{L}{\pi}P^+$ for total longitudinal momentum $P^+$.
Also $x\equiv p^+/P^+$ becomes $n/K$, with $n$ odd for fermions and 
even for bosons.  The light-cone Hamiltonian
$H_{\rm LC}\equiv P^+P^-$ is independent of $L$.  

Because each $n$ is positive, DLCQ
automatically limits the number of particles to no more than $\sim K/2$.
The integers $n_x$ and $n_y$ range between limits associated with
some maximum integer $N_\perp$ fixed by the invariant-mass cutoff
$m_i^2+p_{\perp i}^2\leq x_i\Lambda^2$ for each particle $i$.  
We then have a finite matrix representation where 
integrals are replaced by discrete sums
\be 
\int dp^+ \int d^2p_\perp f(p^+,\vec{p}_\perp)\simeq
   \frac{2\pi}{L}\left(\frac{\pi}{L_\perp}\right)^2
   \sum_{\underline{n}}
   f(n\pi/L,\vec{n}_\perp\pi/L_\perp)\,. 
\ee
This trapezoidal approximation is improved through the
inclusion of weighting factors~\cite{PV1} that take into account
distances between grid point locations and the integration
boundaries set by the cutoff $\Lambda^2$.
In the following sections we discuss how these techniques
can be applied to Yukawa theory.

\section{Yukawa theory}

The DLCQ Hamiltonian for Yukawa theory, when 
truncated to include only one fermion, is~\cite{McCartorRobertson}
\bea
\lefteqn{H_{\rm LC}=
   \sum_{\underline{n},s}
      \frac{M^2+\delta M^2+(\vec{n}_\perp \pi/L_\perp)^2}{n/K}
          b_{\underline{n},s}^\dagger b_{\underline{n},s}
   +\sum_{\underline{m}i}
          \frac{\mu_i^2+(\vec{m}_\perp \pi/L_\perp)^2}{m/K}
              a_{i\underline{m}}^\dagger a_{i\underline{m}}} \nonumber \\
   && +\frac{g\sqrt{\pi}}{2L_\perp^2}
          \sum_{\underline{n}\underline{m}}\sum_{si}\frac{\xi_i}{\sqrt{m}}
     \left(\left[\frac{\vec{\epsilon}_{-2s}^{\,*}\cdot\vec{n}_\perp}{n/K}
         +\frac{\vec{\epsilon}_{2s}\cdot(\vec{n}_\perp+\vec{m}_\perp)}
                                               {(n+m)/K}        \right]
     b_{\underline{n}+\underline{m},-s}^\dagger b_{\underline{n},s} 
                                            a_{i\underline{m}} 
          + \mbox{h.c.}\right) \nonumber\\
   && +\frac{Mg}{\sqrt{8\pi}L_\perp}
             \sum_{\underline{n}\underline{m}}\sum_{si}\frac{\xi_i}{\sqrt{m}}
     \left(\left[\frac{1}{n/K}+\frac{1}{(n+m)/K}\right]
     b_{\underline{n}+\underline{m},s}^\dagger b_{\underline{n},s} 
                a_{i\underline{m}}    + \mbox{h.c.} \right)  \\
   && +\frac{g^2}{8\pi L_\perp^2}
              \sum_{\underline{n}\underline{m}\underline{m}'}\sum_{sij}
         \frac{\xi_i\xi_j}{\sqrt{mm'}}
     \left[\left(b_{\underline{n}+\underline{m}+\underline{m}',s}^\dagger 
         b_{\underline{n},s} a_{i\underline{m}'} a_{j\underline{m}}
                  \frac{1}{(n+m)/K}+ \mbox{h.c.} \right)\right. \nonumber \\
    &&\rule{0.25in}{0mm} \left.
        + b_{\underline{n}+\underline{m}-\underline{m}',s}^\dagger 
            b_{\underline{n},s} a_{i\underline{m}'}^\dagger a_{j\underline{m}}
                  \left(\frac{1}{(n-m')/K}+\frac{1}{(n+m)/K}\right)\right]\,,
  \nonumber
\eea
where $M$ is the fermion mass, $\mu\equiv\mu_0$ the physical boson mass, 
$\mu_i$ the ith PV boson mass,
$\vec{\epsilon}_\lambda=-\frac{1}{\sqrt{2}}(\lambda,i)$ the
polarization vector for helicity $\lambda$, and
\be 
\left[a_{i\underline{m}},a_{j\underline{m}'}^\dagger\right]
          =\delta_{ij}\delta_{\underline{m},\underline{m}'}\,, \;\;
   \left\{b_{\underline{n},s},b_{\underline{n}',s'}^\dagger\right\}
     =\delta_{\underline{n},\underline{n}'} \delta_{s,s'}\,.   
\ee
Modes with zero longitudinal momentum have not been included.  
Fermion self-induced inertia terms are also not
included, because they cancel between PV-boson terms.
A fermion mass counterterm has been included to 
remove shifts proportional to $\ln\mu_i/\mu$.

The number of PV flavors is three.  Their couplings are given
by $\xi_ig$, where $\xi_i=\sqrt{|C_i|}$ and
\be 
1+\sum_{i=1}^3 C_i=0\,, \;\;
\mu^2+\sum_{i=1}^3 C_i\mu_i^2=0\,, \;\;
\sum_{i=1}^3 C_i\mu_i^2\ln(\mu_i^2/\mu^2)=0\,.
\ee
The sign of $C_i$ determines the norm.  This arrangement is
known~\cite{ChangYan,PV1} to produce the cancellations needed
to regulate the theory and restore chiral invariance in the
$M=0$ limit.

To $H_{\rm LC}$ we must add an effective interaction,
modeled on the missing fermion Z graph,
to accomplish cancellation of an infrared singularity in
the instantaneous fermion interaction.  The singularity
occurs when the longitudinal momentum of the instantaneous
fermion approaches zero.  The effective interaction
is constructed from the pair creation
and annihilation terms in the Yukawa light-cone energy 
operator~\cite{McCartorRobertson}
\bea
{\cal P}_{\rm pair}^-&&= \frac{g}{2L_\perp\sqrt{L}}
   \sum_{\underline{p}\underline{q}si}
  \left[\frac{\vec{\epsilon}_{-2s}\cdot\vec{p}_\perp}{p^+\sqrt{q^+}}
        +\frac{\vec{\epsilon}_{2s}^{\,*}\cdot(\vec{q}_\perp-\vec{p}_\perp)}
                    {(q^+-p^+)\sqrt{q^+}}\right]\xi_i
     b_{\underline{p},s}^\dagger d_{\underline{q}-\underline{p},s}^\dagger 
                               a_{i\underline{q}}    +\mbox{h.c.} \\
   && +\frac{Mg}{2L_\perp\sqrt{2L}}
   \sum_{\underline{p}\underline{q}si}
    \left[\frac{1}{p^+\sqrt{q^+}}-\frac{1}{(q^+-p^+)\sqrt{q^+}}\right]\xi_i
     b_{\underline{p},s}^\dagger d_{\underline{q}-\underline{p},-s}^\dagger 
               a_{i\underline{q}}     +\mbox{h.c.}\,,  \nonumber
\eea
combined with the denominator for the intermediate state
\be 
\frac{M^2}{P^+}-p_{\rm spectators}^- -\frac{M^2+p_\perp^{\prime 2}}{p'^+}
         -\frac{M^2+(\vec{q}_\perp^{\,\prime}-\vec{p}_\perp)^2}{q'^+-p^+}
          -\frac{M^2+p_\perp^2}{p^+}
\,. 
\ee
To complete the cancellation of the singularity in the kinematic
regime of positive longitudinal fermion momentum, the instantaneous
interaction is kept only if the corresponding crossed boson graph 
is permitted by the numerical cutoffs.

The single-fermion eigenstate of the Hamiltonian is written
\bea
\Phi_\sigma&=&\sqrt{16\pi^3P^+}\prod_i\sum_{n_i}
     \int\frac{dp^+d^2p_\perp}{\sqrt{16\pi^3p^+}}
   \prod_i\prod_{j_i=1}^{n_i}
     \int\frac{dq_{j_i}^+d^2q_{\perp {j_i}}}{\sqrt{16\pi^3q_{j_i}^+}} 
        \sum_s\\
   &  & \times \delta(\underline{P}-\underline{p}
                     -\sum_i\sum_{j_i}^{n_i}\underline{q}_{j_i})
     \phi_{\sigma s}^{(n_i)}(\underline{q}_{j_i};\underline{p})
     \frac{1}{\sqrt{\prod_i n_i!}}b_{\underline{p}s}^\dagger
      \prod_i\prod_{j_i}^{n_i} a_{i\underline{q}_{j_i}}^\dagger |0\rangle \,,
\nonumber
\eea
with normalization
\be 
\Phi_\sigma^{\prime\dagger}\cdot\Phi_\sigma
=16\pi^3P^+\delta(\underline{P}'-\underline{P})\,. 
\ee
The wave functions $\phi_{\sigma s}^{(n_i)}$ describe a contribution
with $n_0$ physical bosons and $n_i$ bosons of the ith PV flavor.

Mass renormalization is carried out by fixing
the mass $M^2$ of the dressed single-fermion state.  This
is imposed by rearranging the eigenvalue problem 
$H_{\rm LC}\Phi_\sigma=M^2\Phi_\sigma$ 
into an eigenvalue problem for $\delta M^2$:
\bea
x\left[M^2-
   \frac{M^2+p_\perp^2}{x}\right.
   &-&\left.\sum_i\frac{\mu_i^2+q_{\perp i}^2}{y_i}\right] \tilde{\phi} 
                                          \nonumber \\
&-&\int\prod_j dy'_j d^2q'_{\perp j}\sqrt{xx'}{\cal K}\tilde{\phi}'=
                             \delta M^2\tilde{\phi}\,,
\eea
where the new wave functions are related to the originals
by $\tilde{\phi}=\phi/\sqrt{x}$
and ${\cal K}$ is the interaction kernel. 

To fix the coupling we set a value for the expectation value
$\langle :\!\!\phi^2(0)\!\!:\rangle
\equiv\Phi_\sigma^\dagger\!:\!\!\phi^2(0)\!\!:\!\Phi_\sigma$
for the boson field operator $\phi$.
From a numerical solution it can be computed fairly efficiently in a sum
similar to a normalization sum
\bea
\langle :\!\!\phi^2(0)\!\!:\rangle
        &=&\prod_i\sum_{n_i=0}\prod_i\prod_{j_i}^{n_i}
                          \int\,dq_{j_i}^+d^2q_{\perp {j_i}} \sum_s
   \\
  & & \rule{0.25in}{0mm} \times \left(\sum_{k=1}^n \frac{2}{q_k^+/P^+}\right)
              \left|\phi_{\sigma s}^{(n_i)}(\underline{q}_{j_i};
       \underline{P}-\prod_i\sum_{j_i}\underline{q}_{j_i})\right|^2\,.
\nonumber
\eea
The constraint on $\langle :\!\!\phi^2(0)\!\!:\rangle$ is satisfied by 
solving it simultaneously with the eigenvalue problem.

\section{Diagonalization method}

The method used for diagonalization of the eigenvalue problem
is based on a special form of the Lanczos algorithm~\cite{Lanczos}
designed to handle the indefinite norm.  Let $\eta$ represent
the metric signature, so that numerical dot products are written
$\langle\phi'|\phi\rangle=\vec{\phi}^{\,\prime *}\cdot\eta\vec{\phi}$.
An operator $H$ is self-adjoint with respect to this metric if~\cite{Pauli}
$\bar{H}\equiv\eta^{-1}H^\dagger\eta=H$.  The Lanczos algorithm for
the diagonalization of $H$ then takes the form
\bea 
\vec{q}_{j+1}&=&\vec{r}_j/\beta_j\,, \;\;
   \vec{r}_j=H\vec{q}_j-\gamma_{j-1}\vec{q}_{j-1}-\alpha_j\vec{q}_j 
\nonumber \\
\nu_{j+1}&=&\mbox{sign}(\vec{r}_j^{\,*}\cdot\eta\vec{r}_j)\,,\;\;
   \nu_1=\mbox{sign}(\vec{q}_1^{\,*}\cdot\eta\vec{q}_1)\,, 
\\
\alpha_j&=&\nu_j\vec{q}_j^{\,*}\cdot\eta H\vec{q}_j \,, \;\;
   \beta_j=+\sqrt{|\vec{r}_j^{\,*}\cdot\eta\vec{r}_j|}\,, \;\;
   \gamma_j=\nu_{j+1}\nu_j\beta_j\,.
\nonumber
\eea
The original (large) matrix $H$ acquires a new (small) matrix 
representation $T$ with respect to the basis formed by the 
vectors $\vec{q}_j$:
\be 
H\rightarrow T\equiv\left(\begin{array}{llllll}
              \alpha_1 & \beta_1 & 0 & 0 & 0 & \ldots \\
              \gamma_1 & \alpha_2 & \beta_2 & 0 & 0 & \ldots \\
                0 & \gamma_2 & \alpha_3 & \beta_3 & 0 & \ldots \\
                0 & 0 & \gamma_3 & .  & . & \ldots \\
                0 & 0 & 0 & . & . & \ldots \\
                . & . & . & . & . & \ldots \end{array} \right)\,. 
\ee
The elements of $T$ are real, and the matrix is self-adjoint 
with respect to the induced metric $\nu$.
We can solve $T\vec{c}_i=\lambda_i\vec{c}_i$ for eigenvalues 
and right eigenvectors and find that 
$H\vec{\phi}_i\simeq\lambda_i\vec{\phi_i}$, with
\be 
\vec{\phi}_i=\sum_k(c_i)_k\vec{q}_k\,,\;\;
   \vec{\phi}_i^*\cdot\eta\vec{\phi}_j=\vec{c}_i^{\,*}\cdot\nu\vec{c}_j\,. 
\ee
Extreme eigenvalues are well approximated after only a few iterations.

Because of the indefinite metric, the physical one-fermion state is 
not necessarily the lowest mass state.  It is instead identified by
the following characteristics: a positive norm, a real eigenvalue, and 
the largest bare fermion probability between 0 and 1.
Each of these characteristics can be computed without constructing 
the full eigenvector, provided one saves the first component of each
Lanczos vector $\vec{q}_j$ to reconstruct the bare fermion probability.        

\section{Future work}

Current work on this formulation of Yukawa theory is focused on
the tuning of DLCQ weighting factors,~\cite{PV1} both in general 
and specifically with respect to the infrared cancellations between 
instantaneous fermion interactions, crossed boson graphs, and the 
effective Z interaction.  The singular interactions make the 
numerical representation more sensitive to the weighting than was the 
case for earlier model calculations.~\cite{PV1,PV2}

Within the present no-pair approximation, we can next consider
two-fermion states.  This would allow consideration of a true
bound state.~\cite{YukawaLFTD}  For the full theory, with fermion pair
creation, we can again consider the one-fermion and two-fermion
sectors.  This will require further analysis of the infinities
of the theory and determination of the appropriate PV particle
types and interactions.

Other theories that could be considered include quantum electrodynamics,
where one might use PV regularization to repeat a calculation
by Hiller and Brodsky~\cite{ae} of the electron's
anomalous moment for large coupling.  Quantum chromodynamics
will require a somewhat different approach; a broken supersymmetric 
formulation, with massive partners playing the role of PV particles
may be the correct route. 

\section*{Acknowledgments}

The work reported here was done in collaboration 
with S.J. Brodsky and G. McCartor and was supported
in part by grants of computing time by the Minnesota
Supercomputing Institute and by the Department of Energy,
contract DE-FG02-98ER41087.

\end{document}